\documentclass[sigconf, nonacm]{acmart}
\usepackage{enumitem}
\usepackage{tikz}
\newcommand\vldbdoi{XX.XX/XXX.XX}
\newcommand\vldbpages{XXX-XXX}
\newcommand\vldbvolume{18}
\newcommand\vldbissue{12}
\newcommand\vldbyear{2025}
\newcommand\vldbauthors{\authors}
\newcommand\vldbtitle{\shorttitle} 
\newcommand\vldbavailabilityurl{}
\newcommand\vldbpagestyle{empty}

\author{Annabelle Warner}
\affiliation{%
  \institution{University of Utah}
}
\email{annabelle.warner@utah.edu}

\author{Andrew McNutt}
\affiliation{%
  \institution{University of Utah}
}
\email{andrew.mcnutt@utah.edu}

\author{Paul Rosen}
\affiliation{%
  \institution{University of Utah}
}
\email{paul.rosen@utah.edu}

\author{El Kindi Rezig}
\affiliation{%
  \institution{University of Utah}
}
\email{elkindi.rezig@utah.edu}

\usepackage{xspace}

\newcommand{\sys}{\textsc{Buckaroo}\xspace}

\newcommand*\circled[1]{\tikz[baseline=(char.base)]{
            \node[shape=circle,draw,inner sep=1pt,fill=black,text=white] (char) {#1};}}

\newcommand*\circledwhite[1]{\tikz[baseline=(char.base)]{
            \node[shape=circle,draw,inner sep=1pt,fill=white,text=black] (char) {#1};}}

\begin{document}
\title{\sys{}: A Direct Manipulation Visual Data Wrangler}


\begin{abstract}

  Preparing datasets—a critical phase known as data wrangling— constitutes the dominant phase of data science development, consuming upwards of 80\% of the total project time. This phase encompasses a myriad of tasks: parsing data, restructuring it for analysis, repairing inaccuracies, merging sources, eliminating duplicates, and ensuring overall data integrity. Traditional approaches, typically through manual coding in languages such as Python or using spreadsheets, are not only laborious but also error-prone. These issues range from missing entries and formatting inconsistencies to data type inaccuracies, all of which can affect the quality of downstream tasks if not properly corrected. To address these challenges, we present \sys, a visualization system to highlight discrepancies in data and enable on-the-spot corrections through direct manipulations of visual objects. \sys (1)~automatically finds ``interesting'' data groups that exhibit anomalies compared to the rest of the groups and recommends them for inspection; (2)~suggests wrangling actions that the user can choose to repair the anomalies; and (3)~allows users to visually manipulate their data by displaying the effects of their wrangling actions and offering the ability to undo or redo these actions, which supports the iterative nature of data wrangling.

\end{abstract}

\maketitle

\pagestyle{\vldbpagestyle}
\begingroup\small\noindent\raggedright\textbf{PVLDB Reference Format:}\\
\vldbauthors. \vldbtitle. PVLDB, \vldbvolume(\vldbissue): \vldbpages, \vldbyear.\\
\href{https://doi.org/\vldbdoi}{doi:\vldbdoi}
\endgroup
\begingroup
\renewcommand\thefootnote{}\footnote{\noindent
  This work is licensed under the Creative Commons BY-NC-ND 4.0 International License. Visit \url{https://creativecommons.org/licenses/by-nc-nd/4.0/} to view a copy of this license. For any use beyond those covered by this license, obtain permission by emailing \href{mailto:info@vldb.org}{info@vldb.org}. Copyright is held by the owner/author(s). Publication rights licensed to the VLDB Endowment. \\
  \raggedright Proceedings of the VLDB Endowment, Vol. \vldbvolume, No. \vldbissue\ %
  ISSN 2150-8097. \\
  \href{https://doi.org/\vldbdoi}{doi:\vldbdoi} \\
}\addtocounter{footnote}{-1}\endgroup

\ifdefempty{\vldbavailabilityurl}{}{
  \vspace{.3cm}
  \begingroup\small\noindent\raggedright\textbf{PVLDB Artifact Availability:}\\
  The source code, data, and/or other artifacts have been made available at \url{\vldbavailabilityurl}.
  \endgroup
}

\section{Introduction}

\begin{figure}[!bh]
  \centering
  \includegraphics[width=\linewidth]{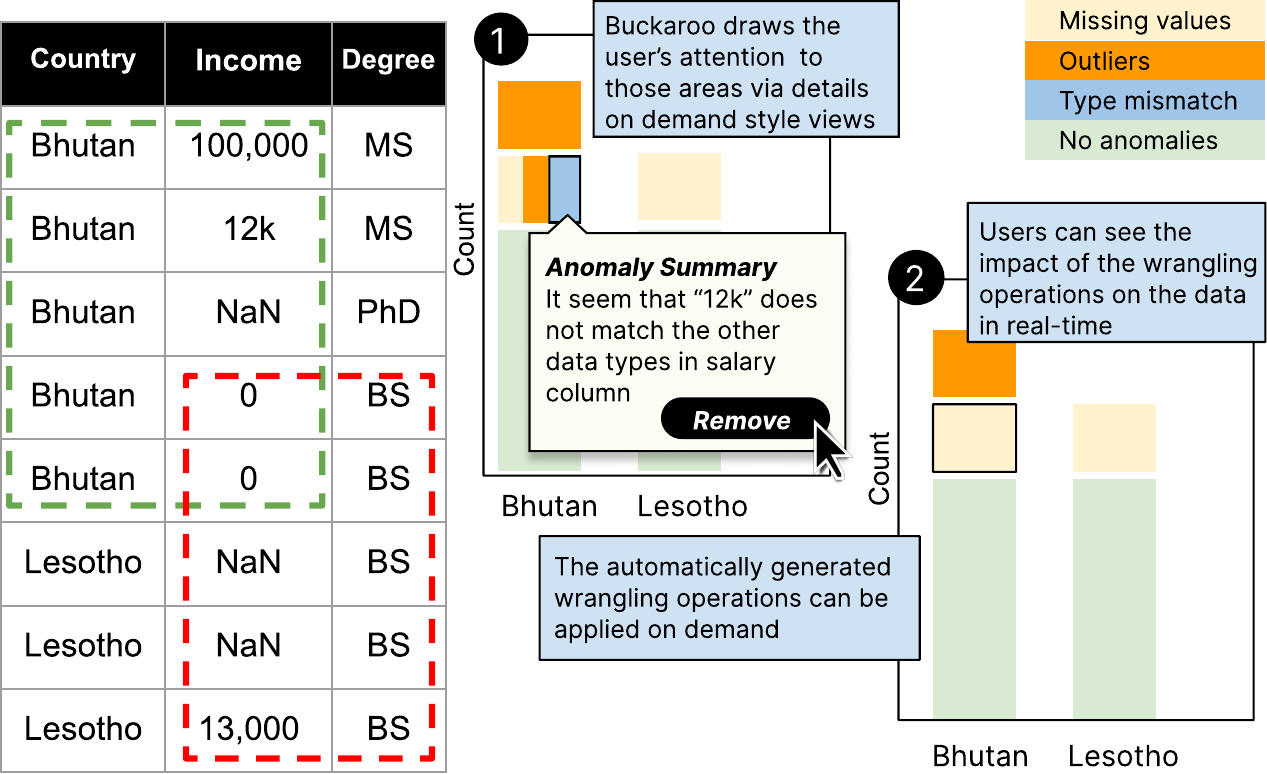}
  \caption{Motivating example: Both country groups (Bhutan and Lesotho) have anomalies such as outliers, missing values, and type mismatches. How to resolve these in an accessible way for non-technical users?
    \sys visually highlights group anomalies and offers corresponding recommended repair operations. The user can then iteratively adjust their data visually until it is reached a satisfactory state.
  }
  \label{fig:example_table}
\end{figure}

Data is ubiquitous, but it is also messy. Before data can be analyzed and acted upon, it must be prepared in a process commonly referred to as data wrangling. However, data wrangling is a well-documented bottleneck in the data science pipeline, accounting for as much as 80\% of the time spent on data analysis~\cite{Deng17, cowrangler}. This process consists of multiple steps, including parsing the data (e.g., converting to a spreadsheet or loading into a database), transforming the data into the right set of rows and columns for analysis, cleaning invalid or missing data, integrating multiple data sources, removing duplicate or similar records, validating that the data is complete and correct, and so on~\cite{retclean, cowrangler}.

Data wrangling is time-consuming and error-prone—issues like missing values or formatting inconsistencies are hard to catch and can lead to incorrect results. Fixes are often applied inconsistently and non-reproducibly through spreadsheets or ad-hoc scripts.

\noindent{\textbf{Visualization for data wrangling.}} Visualization is widely recognized as a powerful tool for data analysis, enabling data scientists to identify patterns, trends, and relationships that might otherwise remain hidden in raw data. Interactive visual analytics systems allow users to dynamically manipulate data representations, facilitating exploration and accelerating the process of insight generation~\cite{heer2012}. Successful applications of visualization have touched every area of science, engineering, business, and the humanities by making large and complex datasets more interpretable and actionable.

Visualization is frequently used in an ad-hoc manner during data wrangling to validate intermediate steps, identify anomalies, and check for correctness. Analysts generate summary statistics, histograms, scatterplots, or other graphics to inspect data distributions, detect missing values, find conflicting value definitions, and ensure transformations have been applied correctly~\cite{ruddle2023tasks}.
This process is typically informal and reactive. While these checks can be invaluable, their ad-hoc nature means they may be inconsistently applied, prone to oversight, and can make them difficult to reproduce.
Moreover, because visualization tools are often separate from the wrangling process itself, switching between data transformations and visual validation can lead to a fragmented workflow where errors may slip undetected through the cracks between steps.

\subsubsection*{Example.}
Lou, a data scientist, is assigned to clean the dataset for further analysis as in \autoref{fig:example_table}. Lou imports the dataset in Python and inspects it for inconsistencies. However, Lou faces several challenges due to the dataset's size: (1)~Detecting subgroups with anomalies can be difficult since such errors are usually infrequent. For instance the income in subgroup \texttt{Country = Bhuthan} has a missing value and two entries with 0; (2)~Correcting these anomalies is tricky, as fixes in one subgroup might introduce new issues in others (for example, removing all rows with an income of 0 from subgroup \texttt{Country = Bhuthan} would leave the subgroup \texttt{Degree = BS} with only a single income entry); and (3)~The data cleaning process requires multiple iterations to develop an effective script.









If Lou instead utilizes \sys{} (\autoref{fig:example_table}\circled{1}), they would receive recommendations on which data groups with anomalies to inspect.
\sys enables Lou to choose specific data wrangling actions tailored to address different types of anomalies. Upon selecting an action, Lou can immediately see its impact (visually) on other groups, and check if it leads to new anomalies. This process is iterative, allowing Lou to apply or revert actions as needed. Once satisfied, Lou can use \sys to generate a Python script of the wrangling steps for future use.

Centering the wrangling process in the visualization itself streamlines data preparation. It would reduce the reliance on iterative scripting and allow for more holistic and immediate error correction, enhancing both the efficiency and accuracy of Lou's work. With this approach, Lou could ensure the dataset is optimally prepared for analysis with less effort and fewer iterations.

\subsubsection*{Contributions.} We introduce \sys, an end-to-end system that enables data wrangling of tables by directly manipulating interactive charts. As illustrated in \autoref{fig:example_table}, \sys:
(1) presents anomalous groups in an interactive chart (e.g., scatterplot, histogram, line chart) to visually expose those groups;
(2)~allows users to explore the various anomalies in the chart, and provides a summary of the anomalies in each group;
(3)~presents a list of actions (wrangling operations) to the user upon selecting a group on the chart to fix the anomaly; and
(4)~supports interactive visual data wrangling, so, users can go back and forth and observe the impact of one fix on other groups interactively until no anomalies are found.

\subsubsection*{Related work. }
\sys bridges the gap between two lines of work: \textbf{(1)~Anomaly detection and data cleaning: } Subgroup discovery is a well-established problem in data engineering~\cite{subgroup1, bach2025usingconstraintsdiscoversparse}. This line of work aims to identify interesting and interpretable subgroups within a dataset where the members of these subgroups exhibit given statistical properties that are significantly different from the general population. A related data wrangling demo is CoWrangler~\cite{cowrangler}, where the goal is to recommend wrangling actions for a given dataset. Unlike those approaches, \sys is orthogonal to the anomaly detection/repair that is being used. Therefore, \sys can augment those methods by providing a direct manipulation visual interface. \noindent{\textbf{(2)~Wrangling script visualization: }}A variety of visualization systems have explored visual support for the data wrangling process. \citet{xiong2022revealing} explore approaches for the visualization of wrangling scripts. We take the opposite tact, by interleaving wrangling activities in the visualizations themselves.
Like us, \citet{kandel2011wrangler} synthesize data transformation scripts based on user intent expression in Wrangler (subsequently commercialized as Trifacta)---a strategy \citet{chen2025dango} more recently extend with natural language prompting.
However, their system makes use of programming-by-demonstration as means of intent expression, whereas we center direct manipulation.



\section{System overview}

\begin{figure}[t]
  \centering
  \includegraphics[width=\linewidth]{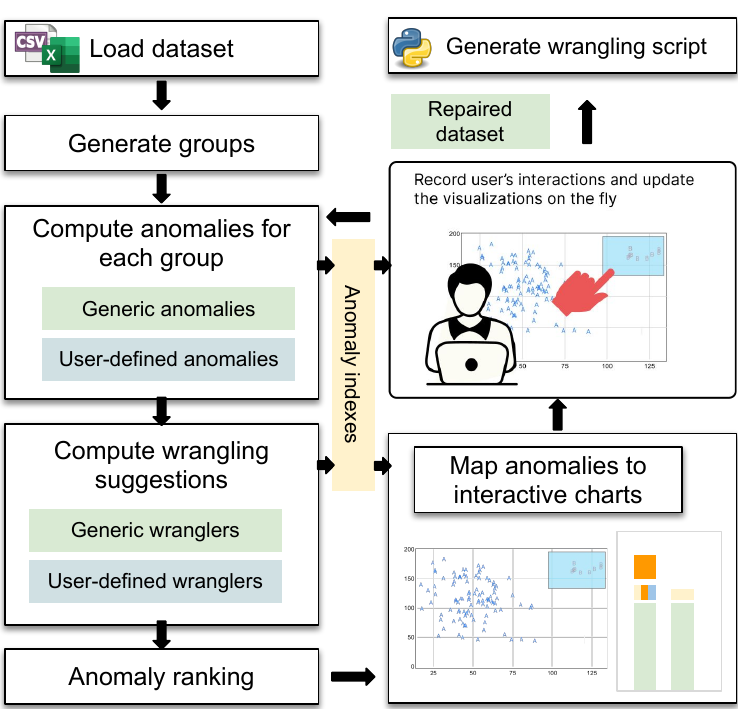}
  \caption{Workflow of \sys: The user uploads a tabular dataset, and \sys generates interactive charts for direct manipulation  anomaly repair.}
  \label{fig:architecture}
  \vspace{-2em}
\end{figure}

The main components of \sys are illustrated in Figure~\ref{fig:architecture}. First, the user uploads a tabular dataset file (e.g., CSV, Excel spreadsheet). Then, \sys generates all the groups formed by categorical attributes. For instance, from Figure~\ref{fig:example_table}, the group \texttt{Country = Bhuthan} is formed by the categorical attribute \texttt{Country}. \sys then computes anomalies and wrangling suggestions for each group, ranks them by importance, and generates the interactive charts. Finally, \sys records the actions taken by the user, and updates the charts when an action is performed.

\subsection{Generating the groups}
\sys generates groups by doing a projection on numerical attributes and grouping by categorical attributes. For instance, in the motivating example (Figure~\ref{fig:example_table}), to get the subgroups containing the income for  \texttt{country = Bhuthan}, we project  \texttt{Income} and group by \texttt{Country}, and then extract the group corresponding to \texttt{Country = ``Bhutan''}. \sys allows users to specify the following:

\begin{itemize}[leftmargin=*]
  \item \textbf{Target attribute: } Users can specify which attributes in the table they are interested in, which would limit the number of generated groups. For instance, in Figure~\ref{fig:example_table}, the target attribute is \texttt{Income}.

  \item \textbf{Granularity specification: } Users can elect to specify the granularity of the generated groups using a
        minimum support parameter, which specifies the minimum number of data points per group.
\end{itemize}

\subsection{Computing group anomalies}

After computing the data groups, \sys proceeds to detect the anomalies. \sys (1)~provides default anomaly types and their detector functions; and (2)~exposes an API that allows users to override these functions or define custom anomaly types with their own detection logic.

\subsubsection*{Default anomaly types}
By default, \sys supports the following data anomalies: (1)~\textbf{Missing values: } Missing values are common in tables. For each data group \sys identifies the missing values; (2)~\textbf{Outliers: } By default, outliers are those values that lie beyond 2 standard deviations from the all-group mean (i.e., a $95\%$ confidence interval); (3)~\textbf{Type mismatch:} This anomaly type is detected when non-numeric values appear in a numeric column; and (4)~\textbf{Group data incompleteness: } Those are groups that have entries lower than a user-specified threshold (2 by default).


\subsubsection*{User-defined anomaly types}
Users often have domain knowledge they want to apply during anomaly detection. For instance, a data scientist analyzing medical data may seek specific anomalies. \sys supports this via an API that lets users define a custom \texttt{Detector} function, which flags anomalous tuples in a target column. \sys treats these functions as black boxes, running them on each data group and collecting the flagged tuples.

\subsubsection*{Indexing anomalies}
To support efficient retrieval, \sys maintains two indexes:
(1) from each anomaly type to the groups exhibiting it, and
(2) from each group to its associated anomaly types.


\subsection{Wrangling suggestions}

After detecting anomalies for each data group, \sys generates wrangling suggestions using default actions or user-defined wranglers.

\subsubsection*{Default wranglers}
\sys provides default actions for common anomalies, which users can override:

\begin{itemize}[leftmargin=*]
  \item \textbf{Missing values:} Suggests imputing with the group/column average or removing the row.
  \item \textbf{Outliers:} Recommends removal or imputation using group/column averages.
  \item \textbf{Type mismatch:} Detects if non-numeric values can be converted (e.g., ``12k'' to 12000), using LLM-generated functions.
  \item \textbf{Group incompleteness:} Suggests merging small groups with similar ones based on the distance between their vector embeddings (e.g., ``USA'' and ``United States of America'').
\end{itemize}

\subsubsection*{User-defined wranglers}
For user-defined anomaly types, users provide an implementation of a \texttt{Wrangler} function for those anomalies. Each anomaly type has a unique identifier, which is then used to map a given wrangling function to an anomaly type.

\subsection{Chart and Code Generation}

\subsubsection*{Anomaly ranking. } Using the indexes built in the anomaly detection step, \sys selects the top-k (k is user-provided)  groups that have the highest number of errors. By default, k is set to 3.


\subsubsection*{Generating the Charts} 
\begin{figure*}[h]
  \centering
  \includegraphics[width=\linewidth]{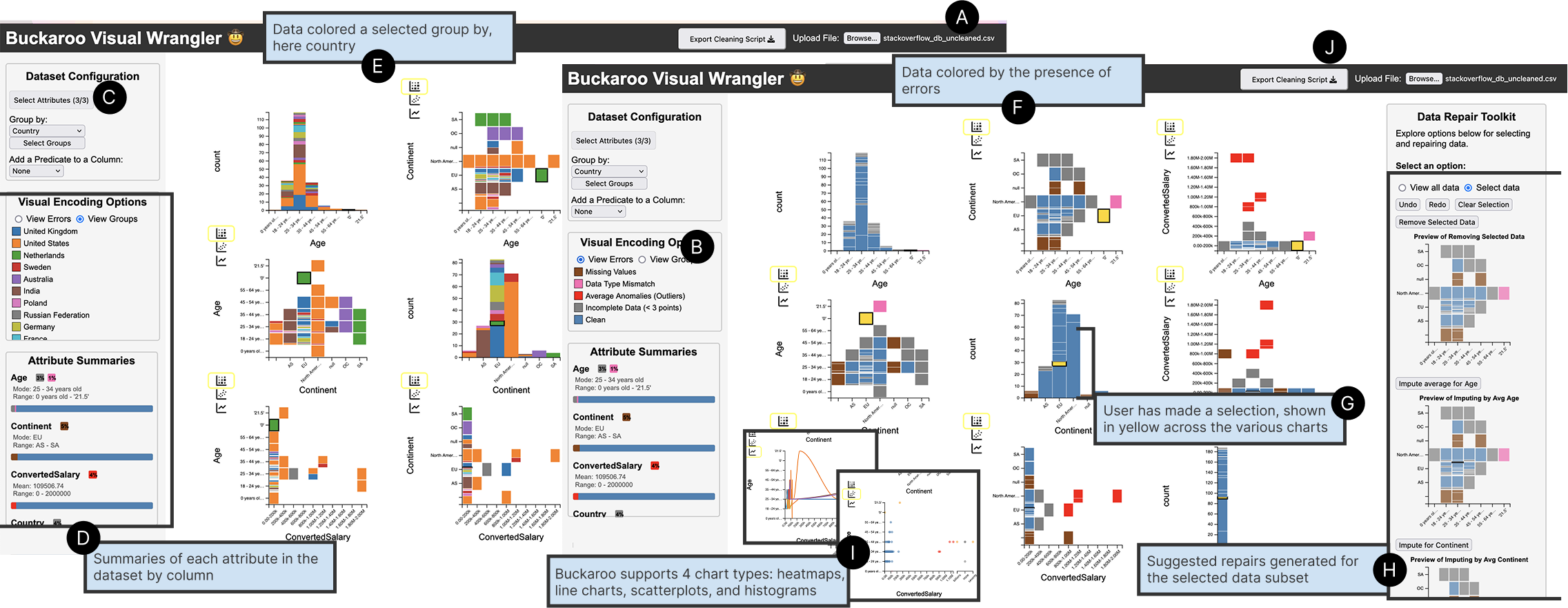}
  \caption{The UI of \sys. \sys features several grouping modes, i.e., by error type~\protect\circled{F} or a group-by attribute~\protect\circled{E}. \sys allows users to visualize errors and perform repairs by suggesting wrangling operations through a repair kit (\protect\circled{H}).}

  \label{fig:screenshot}
  \vspace{-1em}
\end{figure*}

\sys generates a chart matrix (Figure~\ref{fig:screenshot}) where groups are shown in different boxes in the histogram bars. Users can visually inspect various types of group anomalies and directly manipulate them from the charts.
\sys uses the previously built indexes to show the anomaly types per group and the groups per anomaly type. Because it uses anomaly ranking, it can show only the top groups and anomaly types on the charts.

\subsubsection*{Recording the interactions. }
\sys keeps track of all user actions, so they can undo and redo those actions interactively.
This supports the notoriously iterative~\cite{elkindicidr} process of data wrangling.

\subsubsection*{Code generation. } Once the user is satisfied with the wrangling result, \sys generates a stand-alone Python code that loads the input dataset, and includes the sequence of wrangling operations the user applied on the chart.
\sys uses the definitions of the \texttt{Detector} and \texttt{Wrangler} functions to produce this final script.

\newcommand\linesep{0.5em}

\subsubsection*{Implementation} Our demo system is built around a suite of D3-based charts (selected to allow for more bespoke interaction design). We use Arquero~\cite{arquero} for data processing because its straightforward DataFrame like structure scales well on client side tasks (supported by Apache Arrow), and because its small collection of analytic processing verbs have a straightforward mapping to SQL.

\section{Demonstration plan}
We demonstrate \sys on the following real-world datasets: (1)~Stackoverflow survey~\footnote{https://survey.stackoverflow.co}; (2)~Consumer complaints~\footnote{https://www.consumerfinance.gov/data-research/consumer-complaints}; and (3)~Chicago crime~\footnote{https://www.chicago.gov/city/en/dataset/crime.html}. VLDB paricipants will wrangle those datasets to repair their group anomalies visually using \sys. Figure~\ref{fig:screenshot} shows the main user interface of \sys.

\subsubsection*{Demonstration outline.} In this demo we aim to (1)~show the participants how visual data wrangling can significantly make the data wrangling ``fun'' and less prone to errors; (2)~allow the participants to experience the iterative nature of data wrangling through the suggested charts, where the participants can undo and redo actions as needed; and (3)~thanks to its indexes, we show the participants how responsive \sys is in updating the visualizations. In the following, we describe the demonstration scenario steps:

\vspace{\linesep}
\noindent{\textbf{\circledwhite{1}~Uploading the dataset. }} In Figure~\ref{fig:screenshot}~\circled{A}, users start by uploading their dataset file. \sys supports CSV and Excel formats.

\noindent{\textbf{\circledwhite{2}~Setting up anomaly detection. }} After the dataset is uploaded, \sys automatically detects and categorizes data errors, using colors to distinguish error types (Figure~\ref{fig:screenshot}~\circled{B}). \sys adopts a chart matrix to provide users with a comprehensive overview of data errors across multiple combinations of attributes. Users can visualize data errors in two modes:

\noindent{}\textbf{Group name mode:} Using a group-by attribute (Figure~\ref{fig:screenshot}~\circled{C}), users can  view the stacked histograms where the groups are color-coded by values of the group-by column  (Figure~\ref{fig:screenshot}~\circled{E}). Upon hovering on a given group, \sys displays its error types.

\noindent{}\textbf{Error type mode:} In this mode (Figure~\ref{fig:screenshot}~\circled{F}), \sys color-codes the groups (created from the group-by attribute) by their dominant error type.



\noindent{\textbf{\circledwhite{3}~Attribute summary. }} \sys generates a list of recommended attributes to examine (Figure~\ref{fig:screenshot}~\circled{D}). For each attribute, it provides a summary of the associated error types and their frequency relative to the total number of values in that attribute. This feature is crucial to guide users to the most problematic groups.

\vspace{\linesep}
\noindent{\textbf{\circledwhite{4}~Repair kit. }} When the user selects an error bar on the histogram (Figure~\ref{fig:screenshot}~\circled{G}), \sys generates a list of recommended repairs for that error (Figure~\ref{fig:screenshot}~\circled{H}). The impact of each repair on the data is shown visually, allowing the user to assess which option is most appropriate. For example, the user might observe that a certain repair significantly alters the data distribution—typically an undesirable outcome. \sys displays both built-in and user-defined wranglers to assist in repairing the input data.

\vspace{\linesep}
\noindent{\textbf{\circledwhite{5}~Visualization types. }}
\sys supports heatmaps, line charts, scatterplots, and stacked histograms (Figure~\ref{fig:screenshot}~\circled{I}).
These visualizations allow users to explore how different chart types
affect the interpretability of data errors and their suggested repairs.

\vspace{\linesep}
\noindent{\textbf{\circledwhite{6}~Multi-step anomaly detection and wrangling. }} \sys is an interactive system that records all the user's actions,
supporting undo/redo.
We will guide the participants in a multi-step wrangling process where they can try out various repairs and be able to go back and forth until the dataset has been repaired.

\vspace{\linesep}
\noindent{\textbf{\circledwhite{7}~Code generation. }}
Finally, \sys generates a Python script that captures all the wrangling actions applied to the dataset.
This allows users to reapply the same transformations in the dataset for future use  (Figure~\ref{fig:screenshot}~\circled{J}).

\subsubsection*{Demonstration engagement. } In addition to a guided demonstration of \sys, we will encourage participants to upload their own datasets to \sys for a hands-on experience with visual data wrangling. This interactive session aims to highlight how in-visualization data wrangling represents a transformative shift in data preparation, which holds  potential for further exploration.

\bibliographystyle{ACM-Reference-Format}
\bibliography{sample}










\end{document}